\documentclass[aps,prd,twocolumn,superscriptaddress,nofootinbib]{revtex4-2}

\usepackage[T1]{fontenc}
\usepackage{lmodern}
\usepackage{graphicx}

\usepackage{amsmath,amssymb}
\usepackage{booktabs}
\usepackage{float}

\usepackage{xcolor}

\def\sci#1#2{$#1\times10^{#2}$}

\usepackage{hyperref}
\hypersetup{
  colorlinks=true,
  linkcolor=blue,
  citecolor=blue,
  urlcolor=blue
}

\setcitestyle{numbers,sort&compress}

\newcommand{\coh}{T_{\mathrm{coh}}}

\begin{document}

\title{Improvements in Detecting Ultralight Dark Matter with LIGO Data}

\author{Zeyu Xiao}
\affiliation{Department of Physics, University of Michigan, Ann Arbor, Michigan 48109, USA}

\author{Zihao Chen}
\affiliation{Department of Physics, Cornell University, Ithaca, New York 14853, USA}

\author{Keith Riles}
\affiliation{Department of Physics, University of Michigan, Ann Arbor, Michigan 48109, USA}

\date{\today}

\begin{abstract}
We present improved search methods for detecting ultralight dark matter (UDM) in gravitational wave data and apply the techniques to data from the LIGO Hanford (H1) and LIGO Livingston (L1) detectors collected during the first part of the fourth LIGO--Virgo--KAGRA observing run (O4a), targeting both dark photon dark matter (DPDM) and tensor dark matter (TDM). Four improvements with respect to some previously published search methods are used: 1) frequency-dependent coherence times for Fourier transforms; 2) zero-padding to permit use of long coherence times at low frequencies with less sacrifice of livetime; 3) semi-coherent summing of Hanford (H1) and Livingston (L1) detector data; and 4) use of a universal statistic to derive robust upper limits. Together, the first two changes improve sensitivity by allowing more optimal coherence time choices; the latter two changes together also improve sensitivity by allowing robust data combination in the presence of non-Gaussian instrumental artifacts. For the first time, we apply all four of these improvements to UDM searches, using frequency-dependent coherence times of 512\,s, 1024\,s, 2048\,s, 4096\,s, 8192\,s, 16384\,s, and 32768\,s. No outlier is identified as a statistically significant DM candidate, yielding no evidence for either a DPDM or TDM signal in the O4a data. As a result, we obtain the most stringent O4a upper limits to date on the DPDM and TDM coupling strengths over most of the 20--2560\,Hz frequency band searched, corresponding to the UDM mass range \sci{8.3}{-14} to \sci{1.1}{-11}eV$/c^2$.
\end{abstract}

\maketitle
%%%%%%%%%%%%%%%%%%%%%%%%%%%%%%%%%%%%%%%%%%%%%%%%%%%%%%%%%%%%%%%
\section{INTRODUCTION}

The existence of dark matter (DM) is supported by multiple astrophysical and cosmological observations~\citep{Arbey_2021}, yet its nature remains mysterious. In this work, we focus on two ultralight dark matter candidates: dark photon dark matter (DPDM) and tensor dark matter (TDM). Dark photon dark matter corresponds to an ultralight spin-1 vector field that couples weakly to ordinary matter, while tensor dark matter corresponds to an ultralight massive spin-2 field. In the mass range accessible to LIGO, both fields can be treated as coherently oscillating classical backgrounds and can produce narrow-band signals in detector strain data~\citep{PhysRevD.109.095012,LVK_O4a_DM_2025}.

The high displacement sensitivity of LIGO makes it a powerful instrument for searches for such ultralight dark matter signals. An oscillating DPDM field can exert a tiny periodic force on the interferometer mirrors through its weak coupling to baryonic matter, producing an effective strain signal in the detector output ~\citep{Pierce_2018, Guo_2019, Abbott_2022, Abbott_2024_Erratum}. Similarly, tensor dark matter produces a strain signal through an oscillating spin-2 field that induces a differential distortion of the detector arms, analogous to that produced by a gravitational wave~\citep{PhysRevD.109.095012, LVK_O4a_DM_2025}.

A previous search using auxiliary length channels from the KAGRA O3GK run constrained ultralight vector dark matter~\citep{KAGRA_O3GK_2024}. A separate search using O4a data investigated DPDM, TDM, and scalar dark matter, placing upper limits on their corresponding coupling strengths and finding no significant evidence for a signal~\citep{LVK_O4a_DM_2025}. A more recent paper presented methodological improvements, including a frequency-dependent Fourier-transform coherence time for cross-correlation searches and spectral methods to confirm and distinguish possible UDM signals~\cite{Delgado_2026}.

In this paper, we apply four improvements in methodology to search for UDM signals in LIGO Hanford (H1) and LIGO Livingston (L1)~\citep{Aasi_2015} data
collected during the first eight months of the fourth LIGO-Virgo-KAGRA observing run, a period denoted O4a (May 24, 2023 to January 16, 2024). Virgo~\citep{Acernese_2015} did not collect data during O4a, and KAGRA~\citep{Akutsu_2021} collected only one month of data~\citep{O4aOpenData}.  The Advanced LIGO detector performance during O4 and the data quality during O4a are described in Refs.~\citep{Capote_2025, Soni_2025}, respectively.
The four search improvements comprise:
1) frequency-dependent coherence times for ``Short Fourier transforms'' (SFTs);
2) zero-padding to permit use of long coherence times at low frequencies with less sacrifice of livetime;
3) semi-coherent summing of Hanford (H1) and Livingston (L1) detector data; and
4) use of a universal statistic to derive robust upper limits.

An optimum coherence time for UDM searches at a given nominal frequency $f$ is $\sim 10^6/f$~\citep{Pierce_2018}. Requiring contiguous data stretches of at least this coherence time, as has been the norm in previous
UDM searches, leads to substantial livetime loss at low frequencies because of unavoidable interruption of data taking by environmental disturbances. We use coherence times of 512\,s, 1024\,s, 2048\,s, 4096\,s, 8192\,s, 16384\,s, and 32768\,s in roughly octavely distributed frequency bands (with overlap).

Zero-padding between segments allows recovery of otherwise discarded data, improving sensitivity for both the cross-correlation and semi-coherent searches presented here, but with greater gains for cross-correlation because of the previous requirement of simultaneity of contiguous data collection.

In addition, previous use of semi-coherent summing of strain power in UDM searches has relied on coincidence
in outliers between H1 and L1 data to eliminate instrumental artifacts, a coincidence requirement that limits
search sensitivity to that of a single detector. Here we gain in sensitivity by use of semicoherent summing of the independent H1 and L1 strain powers
while relying more heavily upon instrumental line lists to veto outliers~\citep{Goetz_2026}. Even in those bands with vetoed outliers, a universal statistic~\citep{Dergachev_2013} still provides robust upper limits.

In the following, we carry out a search using all of these improvements and find no statistically significant DM candidates. The upper limits obtained confirm an improved sensitivity over nearly the entire band searched with respect to previous DPDM and TDM coupling strengths reported in Ref.~\citep{LVK_O4a_DM_2025}
using the same O4a data set. 

%%%%%%%%%%%%%%%%%%%%%%%%%%%%%%%%%%%%%%%%%%%%%%%%%%%%%
\section{DARK MATTER INTERACTION MODELS}
Since ultralight DM consists of an enormous number of light particles, both DPDM and TDM can be treated as classical oscillating fields. The nominal frequency $f_0$ of this field is fixed by the DM mass:
\begin{equation}
 f_0 = \frac{m_{\rm DM}c^2}{h}.
\end{equation}
Because of the velocity fluctuations (Maxwell--Boltzmann distribution) of dark matter particles in the Galactic halo, there exists a small stochastic frequency spread:
\begin{equation}
\frac{\Delta f}{f_0}
\sim
\frac{v_0^2}{c^2}
\sim
10^{-6},
\end{equation}
where $v_0$ is the virial velocity of dark matter around the center of the Galaxy. This frequency spread implies that the signal has a finite coherence time, during which the signal can be approximately treated as monochromatic. In addition, the finite coherence time implies a finite coherence length, which is much larger than the separation between earth-based interferometers. Therefore, both DPDM and TDM signals are narrow-band, stochastic, and correlated between two detectors.~\citep{Pierce_2018,PhysRevD.109.095012,LVK_O4a_DM_2025}.

\subsection{Dark Photon Dark Matter}
A classical oscillating background field of DPDM would couple to the baryons in the mirrors of LIGO interferometers, causing a differential strain on the detector. The Lagrangian density $\mathcal{L}$ that characterizes the dark photon coupling to a number current density $J^\mu$ of baryons or baryons minus leptons is:
\begin{equation}
\mathcal{L}
=
-\frac{1}{4\mu_0} F^{\mu\nu} F_{\mu\nu}
+\frac{1}{2\mu_0}\left(\frac{m_A c}{\hbar}\right)^2 A^{\mu} A_{\mu}
-e\,\varepsilon\, J^{\mu} A_{\mu},
\end{equation}
where $\mu_0$ is the magnetic permeability in vacuum, $F^{\mu \nu}$ is the electromagnetic field tensor, $m_A$ is the mass of DPDM, $e$ is the electron charge, $A_\mu$ is the dark four-vector potential, and $\varepsilon$ is the strength of the particle/dark photon coupling normalized by the electromagnetic coupling constant~\citep{Abbott_2022}.

There are two effective strains we consider. First, because of the separation of the mirrors in an interferometer, there exists a phase difference in the DPDM field experienced by different mirrors, which will cause a differential change of the arm length in the interferometer. The expression of the corresponding differential strain $h_D$ can be written as~\citep{Pierce_2018,Abbott_2022}:
\begin{align}
\sqrt{\langle h_D^2\rangle}\;=\; C\,\frac{q}{M}\,\frac{v_0}{2\pi c^2}\,\sqrt{\frac{2\rho_{\rm DM}}{\varepsilon_0}}\;\frac{e\,\varepsilon}{f_0}
\end{align}
where $C = \frac{\sqrt{2}}{3}$ is the geometrical factor that comes from averaging all directions of propagation and polarization of DPDM~\citep{Abbott_2022}, $\frac{q}{M}$ is the charge-to-mass ratio of mirrors in LIGO, $v_0 \approx 220\,\mathrm{km/s}$ is the virial velocity~\citep{Smith_2007}, and $\varepsilon_0$ is the permittivity of vacuum space.

The second strain contribution comes from the common motion of the interferometer mirrors under the DPDM background field. Such a common motion will cause an observable signal because of the finite amount of time that light takes to travel between the input and the end mirrors in the interferometer. The corresponding expression of this effective strain can be written as~\citep{Abbott_2022,Morisaki_2021}:
\begin{align}
\sqrt{\langle h_C^2\rangle} \;=\; \tfrac{\sqrt{3}}{2}\;\sqrt{\langle h_D^2\rangle}\;\frac{2\pi f_0 L}{v_0} ,
\end{align}
where $L = 4\,\mathrm{km}$ is the arm length of the gravitational wave interferometer. Combining both contributions, the total effective strain, averaged over time, has the form:
\begin{equation}
\langle h_{\rm tot}^2\rangle
=
\langle h_D^2\rangle
+
\langle h_C^2\rangle.
\end{equation}

\subsection{Tensor Dark Matter}

Tensor dark matter is described by an ultralight massive spin-2 field. Such a field can arise from massive gravity theory and become a dark matter candidate, mixing with a massless graviton~\citep{Marzola_2018}. Unlike DPDM, which causes an observable signal through its coupling to the baryons in the mirrors, TDM causes an observable signal through a stretching of space-time analogous to that caused by conventional, massless gravitational waves. Therefore, the strain induced by TDM can also be measured by LIGO interferometers.

The Lagrangian density that describes the tensor field $\chi_{\mu\nu}$ is~\citep{Armaleo_2020}:
\begin{equation}
\mathcal{L}
=
-\frac{1}{4}
\chi^{\mu\nu}
\mathcal{E}_{\mu\nu,\alpha\beta}
\chi^{\alpha\beta}
-
\frac{1}{8}
\left(
\frac{m_{\rm DM}c}{\hbar}
\right)^2
\left(
\chi_{\mu\nu}\chi^{\mu\nu}
-
\chi^2
\right),
\end{equation}
where $\chi \equiv \chi^{\mu}_{\ \mu}$ and $\mathcal{E}_{\mu\nu,\alpha\beta}$ is a second-order derivative operator~\citep{Aoki_2016}.

The strain on interferometers arises analogously to that from gravitational waves: the ultralight massive spin-2 field causes a differential stretching of space-time in the two interferometer arms. Following the normalization of Ref.~\citep{LVK_O4a_DM_2025}, the corresponding strain can be written as:

\begin{align}
h(f_0)
&=
\frac{2\alpha\Delta\varepsilon}{2\pi f_0 m_{\rm Pl}}
\sqrt{\frac{\rho_{\rm DM}\hbar c}{2}}
\nonumber\\
&\simeq
7.78 \times 10^{-26}
\left(
\frac{\alpha}{5\times10^{-8}}
\right)
\left(
\frac{100\,\mathrm{Hz}}{f_0}
\right),
\end{align}
where
\begin{equation}
\Delta\varepsilon
\equiv
\varepsilon_{ij}
\left(
n^i n^j - m^i m^j 
\right)
\end{equation}
$\boldsymbol{n}$ and $\boldsymbol{m}$ are unit vectors pointing along the two interferometer arms, $\varepsilon_{ij}$ describes each of the five polarizations of the ultralight tensor field, $\alpha$ is the Yukawa coupling strength, and $m_{\rm Pl}$ is the reduced Planck mass. In obtaining the numerical value above, we use $\Delta\varepsilon=\sqrt{\frac{2}{5}}$,
$\rho_{\rm DM}=0.4\,\mathrm{GeV\,cm^{-3}}$,
and the reduced Planck mass.

%%%%%%%%%%%%%%%%%%%%%%%%%%%%%%%%%%%%%%%%%%
\section{Search methodology}
\label{sec:methodology}

In this section, we explain the considerations leading to the analysis improvements
used in this O4a search. We describe constraints due to the chosen SFT coherence time, the motivation
here for zero-padding, the cross-correlation and semi-coherent methods that include
zero padding and the data segmentation design.

\subsection{Coherence time selection and zero-padding}

As our observation time is much longer than the expected DM coherence time, we model the DM field as a
Maxwell-Boltzmann distribution consistent with $\Delta f/f_0\sim10^{-6}$, for which the characteristic
signal coherence time is $\sim10^6/f_0$, which defines a nominal optimum coherence time over which to
compute a single Fourier transform. Many such ``short Fourier transforms'' (SFTs) are computed over
the observation time in constructing detection statistics for signal detection.
Shorter-than-optimal coherence times lead to larger-than-necessary SFT bin sizes, incurring unnecessary noise, while longer
coherence times lead both to spectral leakage into neighboring bins and to reduced signal coherence in any given SFT because of
signal time evolution.

Practical considerations, however, have led previous analyses based on cross-correlation between the
Hanford H1 and Livingston L1 data sets to make a pragmatic single choice of SFT coherence time $\coh$ = 1800 s (optimal for
a signal frequency of about 555 Hz)~\citep{Guo_2019,Abbott_2022}. Those considerations included the ease of producing SFTs and the evaluation
of background expectation, but most importantly, avoiding the loss in available livetime from requiring contiguous
data samples with an optimal coherence time at low frequencies. Since finely tuned gravitational wave detectors
operating on the Earth's surface are subject to many environmental disturbances, including seismic and anthropogenic ground motion,
interruptions to data taking are common. By using only SFTs with long coherence times
created from contiguous data samples, one necessarily loses more and more livetime as coherence times are lengthened.
This loss is especially acute for cross-correlation analyses using {\it simultaneous} H1 and L1 contiguous data samples.

As discussed below, however, the use of zero-padding of time series data allows the use of long-coherence-time SFTs, even
when the data has gaps. Hence
in this analysis we choose frequency-dependent coherence times, spaced in approximate octaves and with some overlap in
frequency for cross-checking results. Table~\ref{tab:coherence} shows these discrete choices, which range from
32,768 s at 20 Hz down to 512 s at 2600 Hz.
Frequency-dependent coherence times were previously employed in a semi-coherent dark photon search~\citep{Miller_2021}.
Previous semi-coherent UDM searches using non-simultaneous H1 and L1 data have also used frequency-dependent coherence times, but with still finer frequency gradation choices~\citep{LVK_O4a_DM_2025},
to better exploit available livetime. In semi-coherent strain power summing, simultaneity is unnecessary.

The zero-padding used here for constructing SFTs simply sets data samples not taken in observing mode to zero, typically at
the starts or ends of time intervals with the desired SFT coherence times. Such padding preserves the frequency binning
used when summing many SFT powers over the full observation span, but at the expense of reduced effective statistical
weight compared to an SFT of the same length for 100\%\ observing mode data. In addition, zero padding introduces
statistical correlations among SFT bins, which make the statistical properties of the derived detection statistics
more complicated to model. In the following, we describe how SFT weighting is incorporated into cross-correlation
and semicoherent analyses in the presence of zero-padding. 

Adopting common notation in both analyses, let
\(X\in\{H,L\}\) label the detector, \(i\) label a coherent data
segment, \(j\) label a frequency bin with central frequency
\(f_j\) and denote the corresponding complex DFT coefficient by
\(\tilde{D}_{X,ij}\).

The DFT-domain noise power measured from the zero-padded SFT is denoted by \(P^{M}_{X,ij}\) and is defined as
\begin{equation}
P^{M}_{X,ij}
=
\left\langle
\left|\tilde{D}_{X,ij}\right|^2
\right\rangle.
\label{eq:common_noise_power}
\end{equation}
The measured noise power \(P^{M}_{X,ij}\) is estimated using a
bias-corrected running median over a 101-bin window centered on
frequency bin \(j\), including bin \(j\) and 50 bins on each side,
under the assumption of locally flat noise.

For a given segment \(i\), let \(r_{z,X,i}\) denote the fraction of
zero-padded samples in detector \(X\), and define the corresponding
livetime fraction as
\begin{equation}
w_{X,i}
=
1-r_{z,X,i}.
\label{eq:common_zp_weight}
\end{equation}
These definitions are used in both the cross-correlation and semi-coherent analyses below.

\subsection{Cross-correlation Method}
For each simultaneous H1--L1 segment pair, the inverse-variance-optimal
cross-correlation weight at frequency bin \(j\), derived in Appendix~\ref{app:optimal_weighting}, is
\begin{equation}
W^{\mathrm{opt}}_{{\rm CC},i}(f_j)
=
\frac{
2\left(w_{H,i}w_{L,i}\right)^2
}{
P^M_{H,ij}P^M_{L,ij}
}.
\label{eq:cc_opt_weight}
\end{equation}
The cross-correlated signal-strength statistic is then defined as
\begin{equation}
S_j
=
\frac{
\displaystyle
\sum_{i=1}^{N_{\mathrm{DFT}}}
W^{\mathrm{opt}}_{{\rm CC},i}(f_j)
\frac{
\tilde{D}_{H,ij}\tilde{D}_{L,ij}^{*}
}{
w_{H,i}w_{L,i}
}
}{
\displaystyle
\sum_{i=1}^{N_{\mathrm{DFT}}}
W^{\mathrm{opt}}_{{\rm CC},i}(f_j)
}
\, .
\label{eq:cc_stat}
\end{equation}
When no zero padding is present, \(w_{H,i}=w_{L,i}=1\) for every
segment, and Eq.~\eqref{eq:cc_stat} reduces to the standard
inverse-noise-weighted cross-correlation statistic~\cite{Pierce_2018}.

The extracted mean squared strain in frequency bin $j$,
before applying the model-dependent detector response correction,
is then given by
\begin{equation}
\widehat{\langle h^2\rangle}_{\mathrm{CC},j}
=
-\frac{2}{T_{\mathrm{coh}}^2}
\operatorname{Re}(S_j),
\label{eq:cc_strain}
\end{equation}
where the negative sign reflects the negative overlap reduction functions for the DM signals considered here.

In the absence of a signal, \(S_j\) has zero expectation value under
the assumption of uncorrelated detector noise. With the assumption
of Gaussian detector noise, the variance of the real and imaginary
parts of \(S_j\) is
\begin{equation}
\sigma_j^2
=
\frac{1}{
2\displaystyle
\sum_{i=1}^{N_{\mathrm{DFT}}}
\frac{
\left(w_{H,i}w_{L,i}\right)^2
}{
P^M_{H,ij}P^M_{L,ij}
}
}.
\label{eq:cc_var}
\end{equation}
The frequency- and segment-dependent noise powers in Eq.~\eqref{eq:cc_var}
account for variations in detector sensitivity over the observation time.

The signal-to-noise ratio is then defined as
\begin{equation}
\mathrm{SNR}_j
\equiv
\frac{S_j}{\sigma_j} \, .
\label{eq:cc_snr}
\end{equation}

Because the coherence lengths of both DPDM and TDM are much larger than the separation between detectors H1 and L1, the phase difference of the DM field between the two detectors is approximately zero, so the cross-correlation signal is expected to be predominantly real. In addition, for the DPDM and TDM responses, the relevant overlap reduction functions are negative, so the expected signal contribution appears in the negative real part of the
SNR~\citep{Abbott_2022,PhysRevD.109.095012}.

\begin{table*}
\centering
\begin{tabular}{|c|c|c|}
\hline
\textbf{Coherence Time $T_{coh}$}
&
\textbf{Target Frequency}
&
\textbf{Target Frequency Range}
\\
\hline
32768 s & 30.5 Hz   & [20,40] Hz     \\
16384 s & 61.0 Hz   & [40,80] Hz     \\
8192 s  & 122.0 Hz  & [80,160] Hz    \\
4096 s  & 244.0 Hz  & [160,320] Hz   \\
2048 s  & 488.0 Hz  & [320,640] Hz   \\
1024 s  & 977.0 Hz  & [640,1280] Hz  \\
512 s   & 1953.0 Hz & [1280,2560] Hz \\
\hline
\end{tabular}
\caption{Choices of coherence time $T_{coh}$ and the corresponding
target frequency ranges. The internal analysis ranges are extended
beyond these target ranges to provide overlap between adjacent
coherence-time analyses and enable consistency checks near their
boundaries. The corresponding internal analysis ranges are
$20\text{--}80\,\mathrm{Hz}$, $20\text{--}120\,\mathrm{Hz}$,
$40\text{--}200\,\mathrm{Hz}$, $120\text{--}360\,\mathrm{Hz}$,
$280\text{--}680\,\mathrm{Hz}$, $600\text{--}1320\,\mathrm{Hz}$,
and $1240\text{--}2600\,\mathrm{Hz}$ for the seven coherence-time
choices, respectively.}
\label{tab:coherence}
\end{table*}

\subsection{Semi-Coherent Method}
For detector \(X\in\{H,L\}\), the inverse-variance-optimal weight
for coherent segment \(i\) and frequency bin \(j\), derived in Appendix~\ref{app:optimal_weighting}, is
\begin{equation}
W^{\mathrm{opt}}_{X,i}(f_j)
=
\frac{
w_{X,i}^4
}{
\left(P^M_{X,ij}\right)^2
}.
\label{eq:pf_opt_weight}
\end{equation}
The semi-coherent detection statistic for detector $X$ is then defined as
\begin{equation}
R_{\mathrm{PF}}^{X}(f_j)
=
\frac{
\displaystyle\sum_i
W^{\mathrm{opt}}_{X,i}(f_j)
\frac{P_{X,ij}^{\mathrm{PSD}}}{w_{X,i}}
}{
\displaystyle\sum_i
W^{\mathrm{opt}}_{X,i}(f_j)
}.
\label{eq:pf_stat}
\end{equation}
where the sums extend over the SFTs available for detector \(X\) at the chosen coherence time \(T_{\mathrm{coh}}\) and $P_{X,ij}^{\mathrm{PSD}}$ is the measured one-sided power spectral density for each SFT, which has the form
\begin{equation}
P_{X,ij}^{\mathrm{PSD}}
=
\frac{2}{T_{\mathrm{coh}}}
\frac{|\tilde{D}_{X,ij}|^2}{w_{X,i}}.
\label{eq:pf_psd}
\end{equation}

When no zero padding is
present, \(w_{X,i}=1\) for every segment, and the detection statistic
reduces to the standard inverse-noise-weighted semi-coherent power sum.

After calculating the optimally weighted detection statistic for each detector, we combine the H1 and L1 results using
\begin{equation}
R_{\mathrm{PF}}^{HL}(f_j)
=
\frac{
W_H^{\rm eff}(f_j)R_{\mathrm{PF}}^{H}(f_j)
+
W_L^{\rm eff}(f_j)R_{\mathrm{PF}}^{L}(f_j)
}{
W_H^{\rm eff}(f_j)+W_L^{\rm eff}(f_j)
},
\end{equation}
where
\begin{equation}
W_X^{\rm eff}(f_j)
=
\sum_i
W^{\mathrm{opt}}_{X,i}(f_j)
=
\sum_i
\frac{
w_{X,i}^4
}{
\left(P^M_{X,ij}\right)^2
},
\qquad X\in\{H,L\}.
\label{eq:pf_detector_weight}
\end{equation}

The signal-to-noise ratio (SNR) is then defined as
\begin{equation}
\mathrm{SNR}
=
\frac{
R_{\mathrm{PF}}-\langle R_{\mathrm{PF}}\rangle
}{
\sigma
},
\end{equation}
where \(\langle R_{\mathrm{PF}}\rangle\) is the locally estimated
expectation value of \(R_{\mathrm{PF}}\), and \(\sigma\) is the
corresponding local standard deviation estimated from the same
neighborhood window. Specifically,
\begin{equation}
\begin{aligned}
\langle R_{\mathrm{PF}}\rangle(f_b)
&=
\frac{1}{N_b}
\sum_{\ell\in\mathcal{W}_b}
R_{\mathrm{PF}}(f_\ell),
\\[4pt]
\sigma^2(f_b)
&=
\frac{1}{N_b}
\sum_{\ell\in\mathcal{W}_b}
R_{\mathrm{PF}}^2(f_\ell)
-
\langle R_{\mathrm{PF}}\rangle^2(f_b).
\end{aligned}
\end{equation}

Here
\[
\mathcal{W}_b
=
\left\{
\ell:
|\ell-b|\le H,\,
\ell\neq b,\,
\ell\notin\mathcal{M}
\right\}
\]
is a symmetric neighborhood window of \(\pm H\) bins centered at
\(b\), where \(H\) denotes the half-width of the window in units of
frequency bins. The set excludes the central bin and all bins in the
mask set \(\mathcal{M}\), and \(N_b=|\mathcal{W}_b|\) is the number
of bins included in the local estimate. In this search, we set
\(H=50\), corresponding to a nominal window size \(N_b=2H=100\)
away from band edges.

The extracted mean squared strain for the combined H1–L1 semi-coherent analysis is obtained by subtracting the locally estimated noise background from the detection statistic and multiplying the resulting signal power spectral density by the frequency bin width \(1/T_{\mathrm{coh}}\):
\begin{equation}
\widehat{\langle h^2\rangle}_{\mathrm{SC},j}
=
\frac{1}{T_{\mathrm{coh}}}
\left[
R_{\mathrm{PF}}^{HL}(f_j)
-
\left\langle R_{\mathrm{PF}}^{HL}\right\rangle(f_j)
\right].
\label{eq:pf_strain}
\end{equation}

\subsection{Optimum Segment Selection}
In this search, we optimize the selection of O4a segments at each coherence time by maximizing the summed livetime-product score.
Using the livetime fractions \(w_{H,i}\) and \(w_{L,i}\) defined in
Eq.~\eqref{eq:common_zp_weight}, we define the segment-selection score
\begin{equation}
v_i
=
w_{H,i}w_{L,i}
=
(1-r_{z,H,i})(1-r_{z,L,i}).
\end{equation}
This optimization can be done by dynamic programming techniques~\cite{Bellman_1957,ChenThesis}. Define the total observing time as $T_{\mathrm{obs}}$ (in seconds), and the chosen coherence time as $N_{\mathrm{coh}}$ (in seconds). 

We then define an array $V[j]$ over the integer second indices $0\le j\le T_{\mathrm{obs}}$. For $N_{\mathrm{coh}}\le j\le T_{\mathrm{obs}}$, $V[j]$ represents the segment-selection score $v_i$ of the segment starting at second $j-N_{\mathrm{coh}}+1$ and ending at second $j$. For $0\le j<N_{\mathrm{coh}}$, we set $V[j]=0$ because a segment ending at $j$ cannot span the full coherence time.

Next, we define two additional arrays, $M[j]$ for memory and $T[j]$ for tracking, over the same index range $0\le j\le T_{\mathrm{obs}}$. For indices $0 \le j < N_{\mathrm{coh}}$, we set $M[j]=T[j]=0$. To optimize, we iterate over $N_{\mathrm{coh}} \le j \le T_{\mathrm{obs}}$ and update
$M$ and $T$ based on the following rules:

For each integer second index $j$ in the $M$ array:
\[
M[j] = \max\!\bigl(M[j-1],\, M[j-N_{\mathrm{coh}}] + V[j]\bigr)
\]
Then we define the $T$ array:
\[
\mathrm{T}[j]=
\begin{cases}
0, & \text{if }\mathrm{M}[j]=\mathrm{M}[j-1],\\
1, & \text{if }\mathrm{M}[j]=\mathrm{M}[j-N_{\mathrm{coh}}]+V[j]
\end{cases}
\]

Thus, $M[j]$ records the maximum cumulative segment-selection score up to index $j$, while $T[j]$ captures the optimized segment selection: $T[j]=1$ means that the segment ending at second $j$ is selected. To extract the optimized segment selection, we loop through the $T$
array backward, starting from the last index, with the following steps:
\begin{itemize}
\item Check $T[j]$. If $T[j]=0$, the segment ending at $j$ is not selected, and the next index is $j_{\mathrm{next}} = j-1$.
\item If $T[j]=1$, the segment ending at $j$ is selected, and the next index is $j_{\mathrm{next}} = j-N_{\mathrm{coh}}$.
\end{itemize}
By repeating this process for index $j_{\mathrm{next}}$ until the first index is reached, we can achieve optimal segment selections for all coherence time choices. The optimality of this segment-selection procedure has been checked using computer simulations~\cite{ChenThesis}.

%%%%%%%%%%%%%%%%%%%%%%%%%%%%%%%%%%%%%%%%%%%%%%%%%%%%%%%%
\section{Results}
\subsection{Cross-correlation Method}

Following Ref.~\citep{Abbott_2022}, we adopt
$\mathrm{Re}(\mathrm{SNR})<-5.8$ as the threshold for a potential detection candidate
while also following up any outlier satisfying
$\left|\mathrm{Re}(\mathrm{SNR})\right|>5$ or
$\left|\mathrm{Im}(\mathrm{SNR})\right|>5$
across the seven coherence-time analyses. For each outlier, we also examine the ten frequency-lag control samples for evidence of an elevated or loud lag background. Instrumental-line vetoes below 2000\,Hz were based on the O4a H1 and L1 line lists (LIGO-T2400204-v2)~\citep{O4aLineList}. Above 2000\, Hz, the excluded frequency intervals and their boundaries were adopted from Ref.~\citep{Qiao:2026HighFrequency}. After applying these vetoes and rejecting outliers with frequency-lag backgrounds exceeding a threshold corresponding to one expected Gaussian background excursion within the local $\pm0.1$\,Hz window, two outliers remain, consistent with expectation.

\subsection{Semi-Coherent Method}
For a standard normal random variable $Z$, the
corresponding one-sided tail probability is
$p_5=\Pr(Z>5)=\frac{1}{2}\operatorname{erfc}(5/\sqrt{2})
\simeq2.87\times10^{-7}$. Treating the results from different coherence-time
analyses as distinct trials, including those in overlapping frequency ranges,
gives $8\,642\,560$ nominal trials and hence a nominal expectation of $8\,642\,560\,p_5\simeq2.48$ positive Gaussian-noise outliers. This value serves only as an approximate benchmark because the combined SNR is based on locally estimated background quantities; neighboring frequencies can be correlated, and the detector data contain non-Gaussian spectral artifacts.

To suppress outliers caused by a large disturbance in only one detector,
for outliers satisfying $\mathrm{SNR}_{HL}>5$, we additionally require $\mathrm{SNR}_{H}>3.5$ and $\mathrm{SNR}_{L}>3.5$; the value 3.5 corresponds to $5/\sqrt{2}\simeq3.5$ when the two detectors contribute equally.  Applying these additional single-detector SNR requirements leaves six outliers. Further investigation identifies three as associated with simulated gravitational-wave signals ("hardware injections"~\cite{Baxi_2026}) and one with an obvious instrumental spectral line, leaving two outliers. This remaining count is of the same order as the nominal Gaussian expectation, although the latter does not account for the additional selection criteria and vetoes. Neither of the two remaining outliers is identified as a statistically significant DM candidate.

%%%%%%%%%%%%%%%%%%%%%%%%%%%%%%%%%%%%%%%%%%%%%
\subsection{Upper limits}

Since no surviving candidates are found in either analysis, we set 95\%
confidence-level upper limits on the coupling strengths of both DPDM and TDM. For the cross-correlation method, the number of remaining narrow-band artifacts after the veto procedure is small, and the background distribution is sufficiently consistent with the Gaussian assumption. Therefore, we calculate the upper limits with the Feldman--Cousins approach~\citep{Feldman_1998}, in which the real part of the cross-correlation SNR is treated as a Gaussian-distributed zero-mean statistic in the absence of a signal.

For the semi-coherent method, the situation is different. Although known instrumental lines are vetoed before setting the limits, the combination of the H1 and L1 semi-coherent spectra retains a substantially larger number of narrow-band disturbances such that the background cannot be assumed to be a Gaussian distribution. We therefore calculate the semi-coherent upper limits using the Universal Statistic method of Ref.~\citep{Dergachev_2013}. This method provides statistically valid upper limits without assuming a particular {\it a priori} background distribution, while remaining close to the optimal sensitivity for well-behaved Gaussian or exponential noise.

Because the signal has a finite linewidth and a random alignment with FFT bin boundaries, a per-bin search captures, on average, only a fraction of the total signal power in the loudest bin~\citep{Guo_2019,LVK_O4a_DM_2025}. We account for this effect by introducing an efficiency factor $\eta(f;T_{\mathrm{coh}})$, defined as the mean fraction of signal power captured by the loudest frequency bin. The squared strain upper limit is therefore corrected as $h_{95}^2 \rightarrow h_{95}^2/\eta$ before conversion to a coupling limit. The dependence of $\eta$ on frequency and coherence time is shown in Fig.~\ref{fig:app_efficiency}.

The resulting upper limits on the DPDM coupling strength are compared with those of Ref.~\citep{LVK_O4a_DM_2025} in Fig.~\ref{fig:DM_CC_SC}(a). For DPDM, the cross-correlation method is much less sensitive to the common-motion contribution caused by the finite light travel time in the interferometer arms because the corresponding overlap reduction function between the H1 and L1 interferometers is strongly suppressed. Consequently, the semi-coherent DPDM limits are more stringent than the cross-correlation limits~\citep{Morisaki_2021,PhysRevD.109.095012,LVK_O4a_DM_2025}.

The zero-padding cross-correlation limits are more stringent because zero-padding allows valid data in partially live segments to be retained rather than discarding the entire segment, thereby permitting a more nearly optimal choice of $T_{\mathrm{coh}}$ in each frequency band.

The zero-padding semi-coherent limits are also more stringent than the O4a LVK BSD limits~\cite{LVK_O4a_DM_2025} over most of the analyzed frequency range because zero-padding retains valid data in partially live segments and the direct weighted combination of H1 and L1 powers uses information from both detectors. The limits are similar in sensitivity to those of the LVK LPSD limits~\cite{LVK_O4a_DM_2025}, which also combine H1 and L1 spectral noise estimates. At high frequencies, the upper limits from both zero-padding analyses are fully consistent with the corresponding limits in Ref.~\citep{LVK_O4a_DM_2025}, because the shorter coherence times make individual segments less susceptible to interruptions and therefore reduce the impact of zero-padding. At low frequencies, the longer coherence times make interruptions more consequential, leading to a substantially larger improvement from zero-padding.

The resulting upper limits on the TDM coupling strength are compared with those of Ref.~\citep{LVK_O4a_DM_2025} in Fig.~\ref{fig:DM_CC_SC}(b). For TDM, the tensor field produces a differential strain analogous to that caused by gravitational waves, and this strain is correlated between the two detectors because the coherence length of the TDM field is much larger than the separation between H1 and L1. Therefore, the cross-correlation method can use the correlated strain response between the two detectors while suppressing uncorrelated detector noise, leading to more stringent limits than the semi-coherent method~\citep{PhysRevD.109.095012,LVK_O4a_DM_2025}. The cross-correlation improvement relative to Ref.~\citep{LVK_O4a_DM_2025} results from the recovery of livetime and the use of frequency-dependent coherence times, while the semi-coherent improvement additionally benefits from the direct weighted combination of H1 and L1 powers.

 Overall, the upper-limit results show that the zero-padding technique can be applied consistently to searches for both DPDM and TDM. For the cross-correlation analysis, the Feldman--Cousins construction is appropriate because the surviving background is sufficiently close to Gaussian after line vetoes. For the semi-coherent analysis, the Universal Statistic provides a more robust upper-limit construction in the presence of the larger number of residual spectral artifacts. In both cases, the use of different coherence times allows the analysis to recover sensitivity that would otherwise be reduced by interruptions in the detector livetime and by the use of a fixed coherence time.

 %%%%%%%%%%%%%%%%%%%%%%%%%%%%%%%%%%%%%%
\section{Conclusion}
In this search, we employ a zero-padding technique in both the cross-correlation and semi-coherent methods to search for DPDM and TDM using LIGO O4a data. Zero-padding lets us retain data segments with interruptions in detector livetime, apply appropriate weighting factors, and use optimized coherence times across frequency bands. With the appropriate weighting factors and careful veto procedures, we obtain more stringent upper limits while finding no significant candidates for either DM model. The improvement is particularly significant at low frequencies, where longer coherence times can be used, and the effect of zero-padding is most beneficial. Hence, zero-padding offers an attractive method to achieve improved sensitivity in future UDM searches. 

\begin{acknowledgments}
We thank Mai Qiao for helpful comments on this manuscript. This material is based upon work supported by NSF’s LIGO Laboratory, which is a major facility fully funded by the National Science Foundation. The authors are grateful for computational resources provided by the LIGO Laboratory and supported
by National Science Foundation Grant PHY-2408883. This document has been assigned LIGO document number P2600619.
This research has made use of data or software obtained from the Gravitational Wave Open Science Center (gwosc.org), a service of the LIGO Scientific Collaboration, the Virgo Collaboration, and the KAGRA Collaboration. 
\end{acknowledgments}

\par
\onecolumngrid

\begingroup
\centering

\refstepcounter{figure}
\label{fig:DM_CC_SC}

\includegraphics[width=1.1\textwidth]{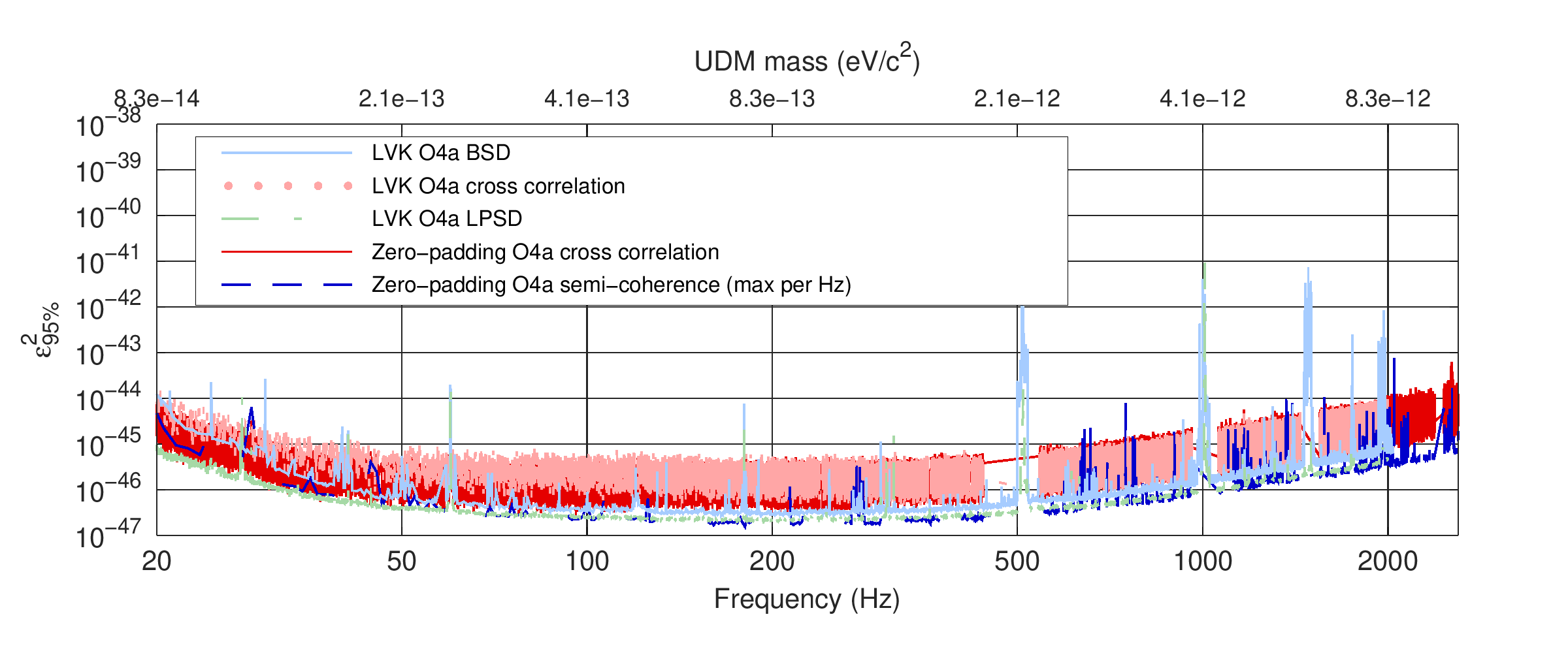}
\vspace{-0.5cm}

{\small (a) Dark photon dark matter\par}

\vspace{0.1cm}

\includegraphics[width=1.1\textwidth]{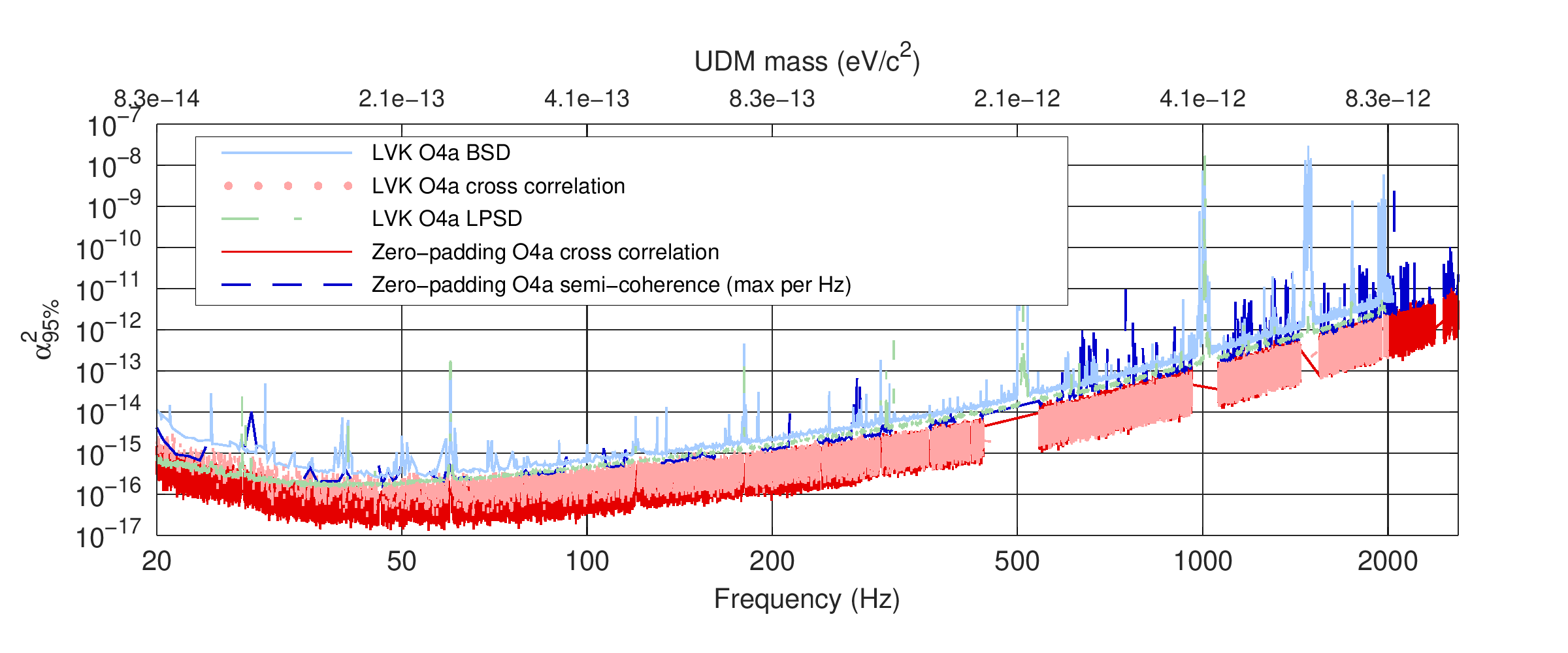}
\vspace{-0.15cm}

{\small (b) Tensor dark matter\par}

\vspace{0.15cm}

\begin{minipage}{\textwidth}
\small
\noindent
\textbf{FIG.~\thefigure.}\quad
Upper limits on the coupling strengths of dark photon dark matter and tensor dark matter obtained with the zero-padding cross-correlation and semi-coherent methods, compared with the O4a cross-correlation, BSD, and LPSD results in Ref.~\citep{LVK_O4a_DM_2025}, using the publicly released upper-limit data~\citep{O4a_DM_DataProduct}.
(a) Upper limits on the square of the dark-photon/baryon coupling
strength, $\varepsilon_{95\%}^{2}$.
(b) Upper limits on the square of the Yukawa coupling strength of
tensor dark matter, $\alpha_{95\%}^{2}$.
For both diagrams, the zero-padding curves use mean values for overlap
bands between different coherence times.
\end{minipage}

\par
\endgroup
\twocolumngrid
%%%%%%%%%%%%%%%%%%%%%%%%%%%%%%%%%%%%%%%%%%%%%%%%%%
\onecolumngrid
\appendix

\section{Supplementary figures}
\vspace{-1cm}
\begin{figure}[H]
\centering
\includegraphics[width=0.8\textwidth]{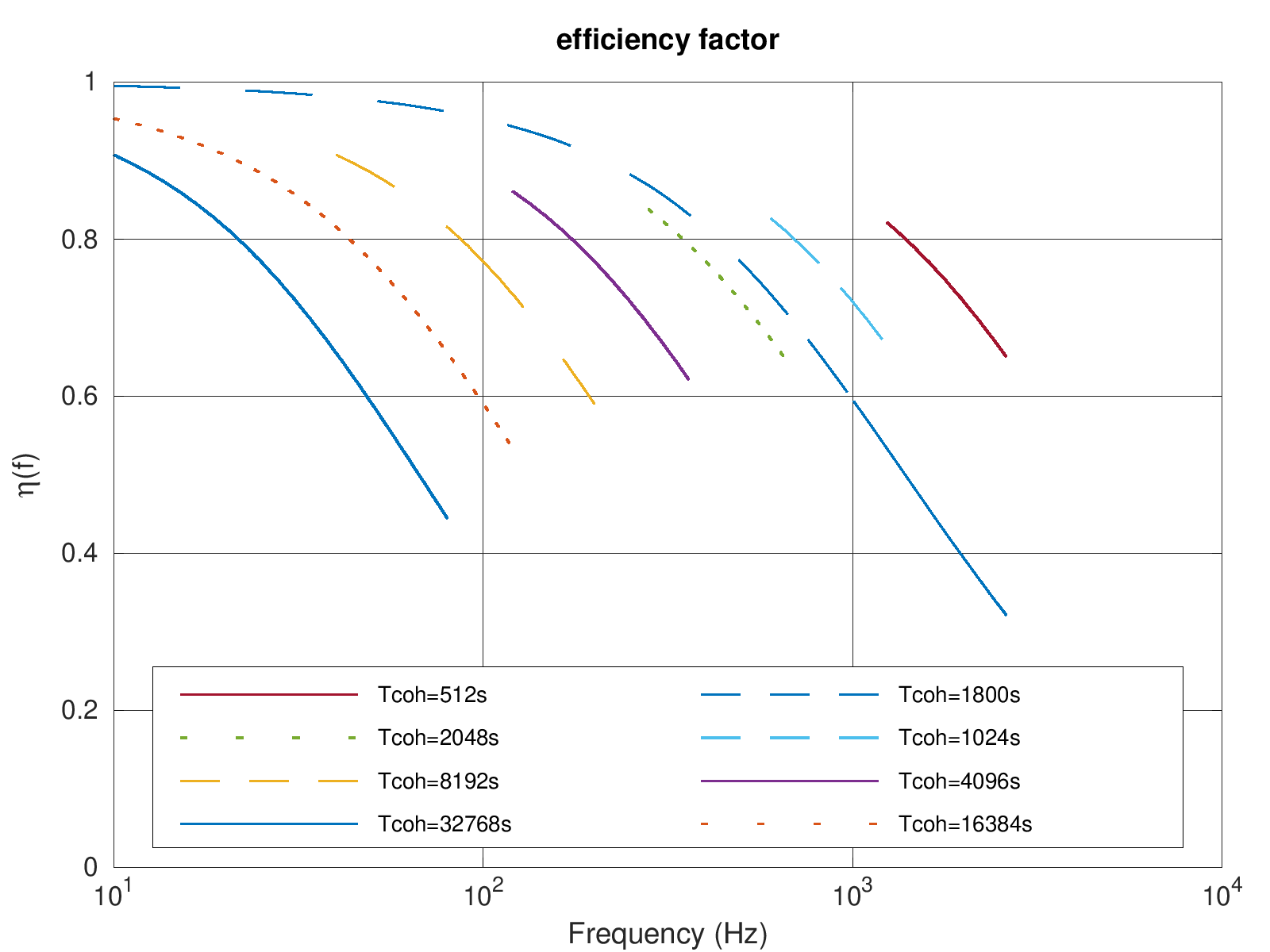}
\caption{
Bin-averaged efficiency factors due to spectral leakage for the different coherence-time selections in their
corresponding target frequency bands.
}
\label{fig:app_efficiency}
\end{figure}

\begin{center}

\includegraphics[width=1.1\textwidth,height=1\textheight,keepaspectratio]{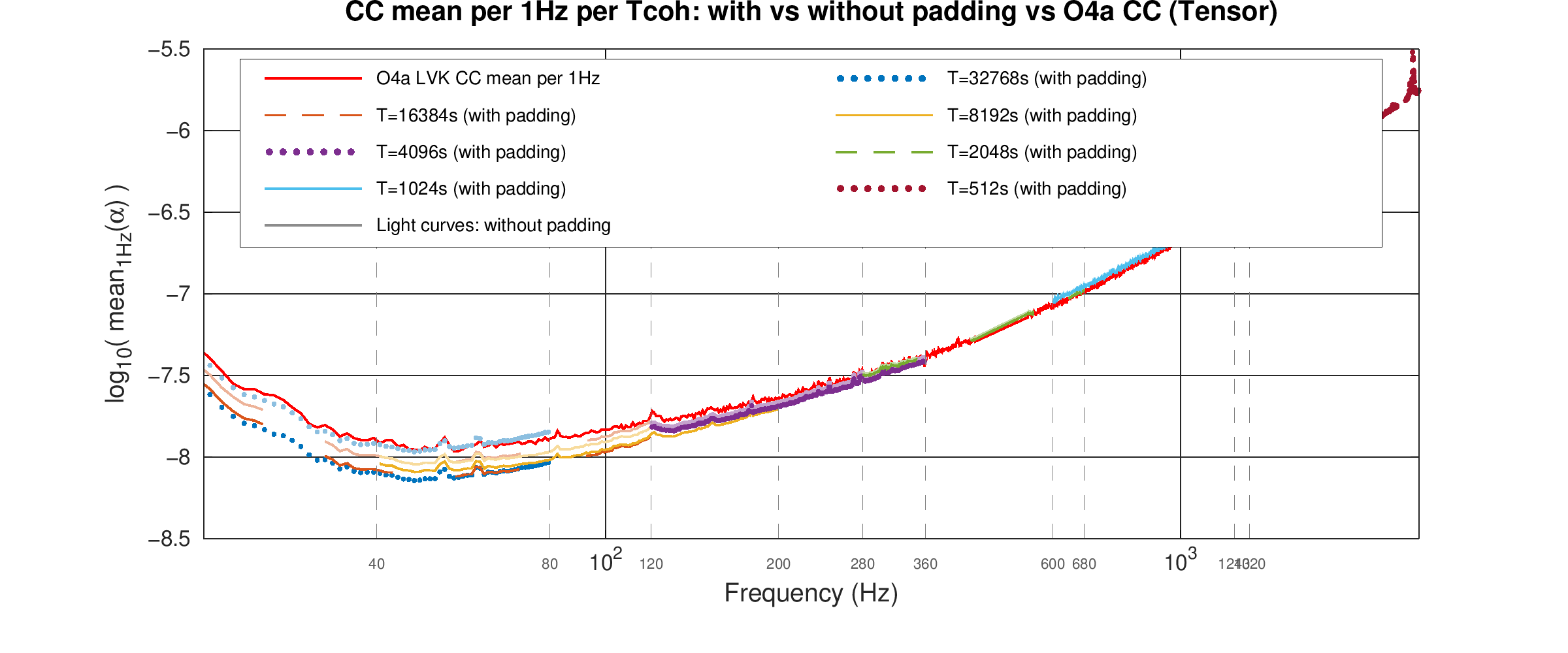}
\vspace{-0.1cm}

{\small (a) Cross-correlation\par}

\vspace{0.15cm}

\includegraphics[width=1.1\textwidth,height=1\textheight,keepaspectratio]{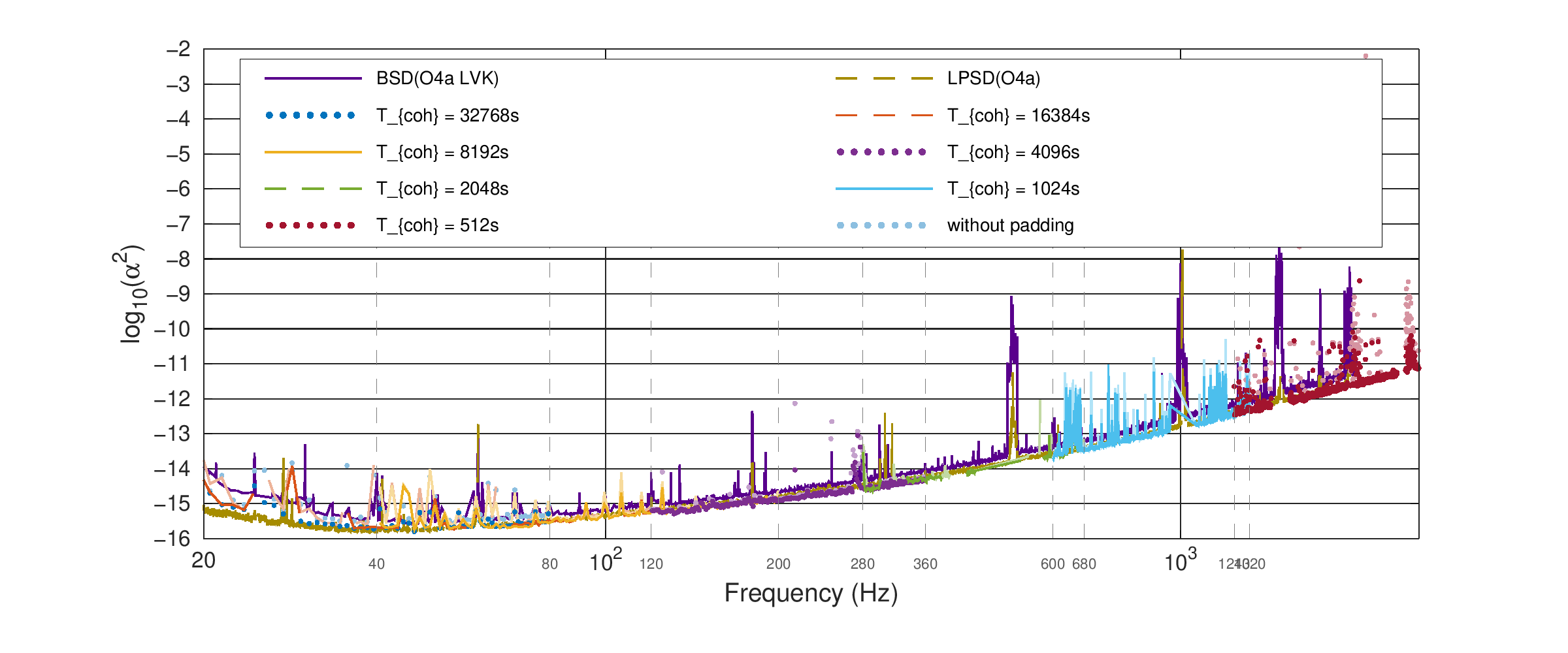}
\vspace{-0.1cm}

{\small (b) Combined H1--L1 semi-coherent\par}

\end{center}

\begingroup
\footnotesize
\refstepcounter{figure}
\label{fig:app_tdm}
\noindent\textsc{Fig.~\thefigure.}\quad
Upper limits of the square of the Yukawa coupling strength of tensor dark matter, $\alpha_{95\%}^{2}$, with different coherence times.
(a) The mean value of \textbf{Cross-correlation}.
(b) Max per Hz of \textbf{Combined H1--L1 semi-coherent}.
The curves in the overlap regions between adjacent coherence time bands are shown separately.
\par
\endgroup

\clearpage

\begin{center}

\includegraphics[width=1.1\textwidth,height=1\textheight,keepaspectratio]{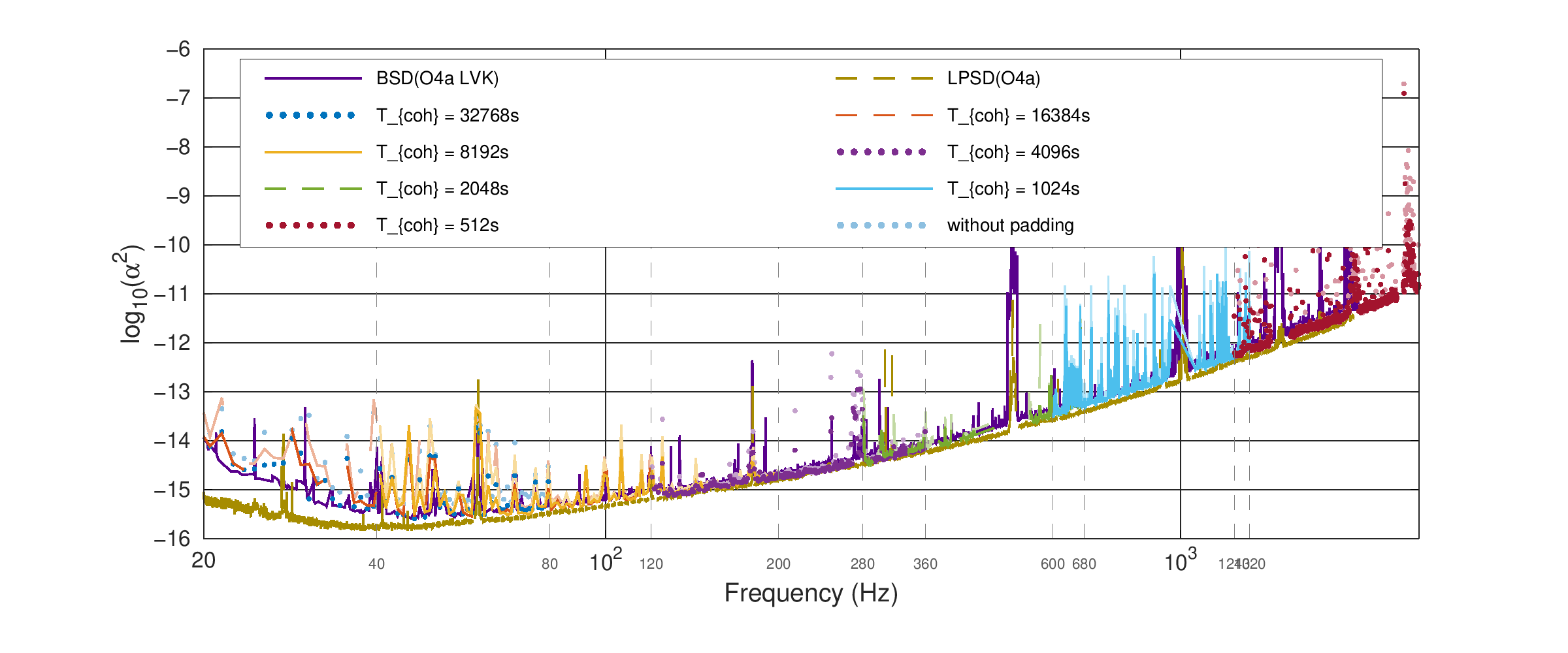}
\vspace{-0.1cm}

{\small (a) H1 semi-coherent\par}

\vspace{0.9cm}

\includegraphics[width=1.1\textwidth,height=1\textheight,keepaspectratio]{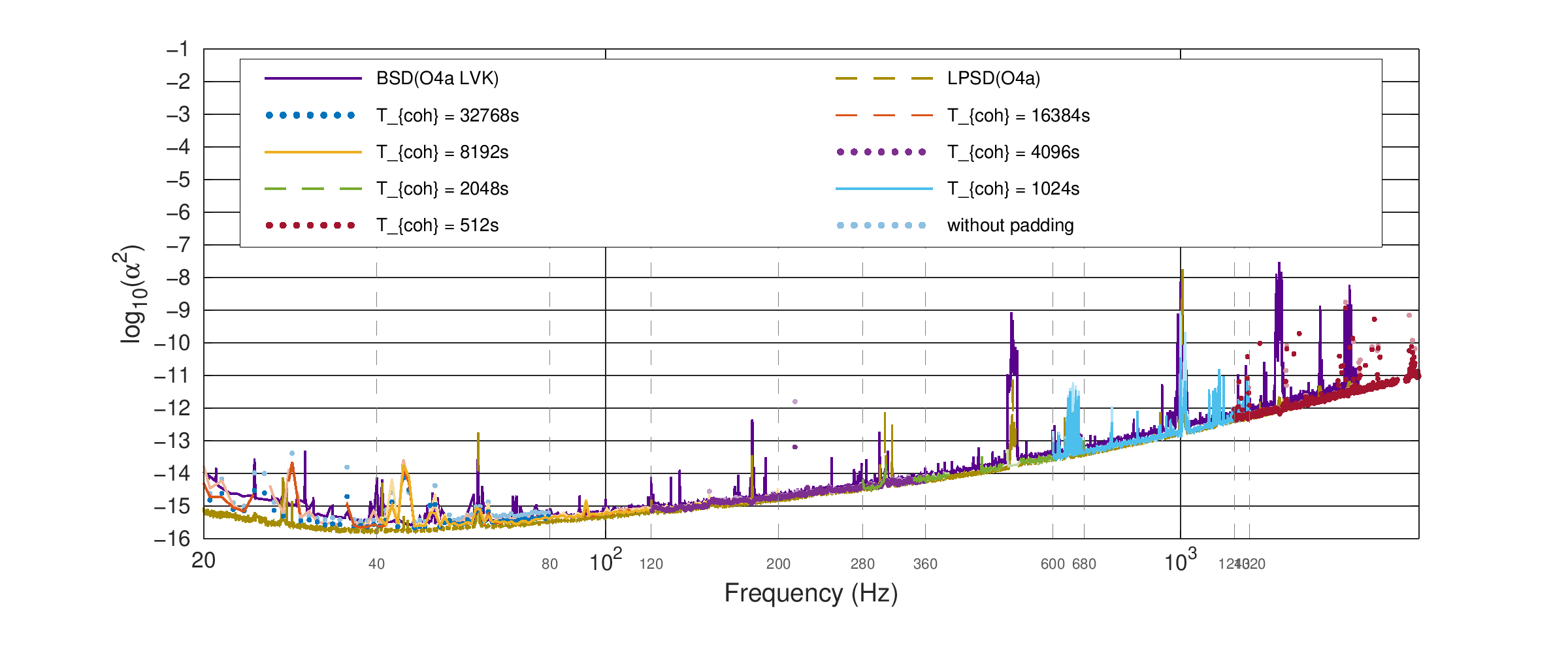}
\vspace{-0.1cm}

{\small (b) L1 semi-coherent\par}

\end{center}

\begingroup
\footnotesize
\refstepcounter{figure}
\label{fig:app_tdm_h1_l1}
\noindent\textsc{Fig.~\thefigure.}\quad
Upper limits of the square of the Yukawa coupling strength of tensor dark
matter, $\alpha_{95\%}^{2}$, with different coherence times.
(a) \textbf{H1 semi-coherent}.
(b) \textbf{L1 semi-coherent}.
The curves in the overlap regions between adjacent coherence-time bands are shown separately. Note that the H1 curve is consistent with the curve of Ref.~\citep{LVK_O4a_DM_2025} at high frequency, while it is weaker at low frequency. This is because the H1 detector is less sensitive than L1 at low frequencies.
\par
\endgroup

\clearpage

\begin{center}

\includegraphics[width=1.1\textwidth,height=1\textheight,keepaspectratio]{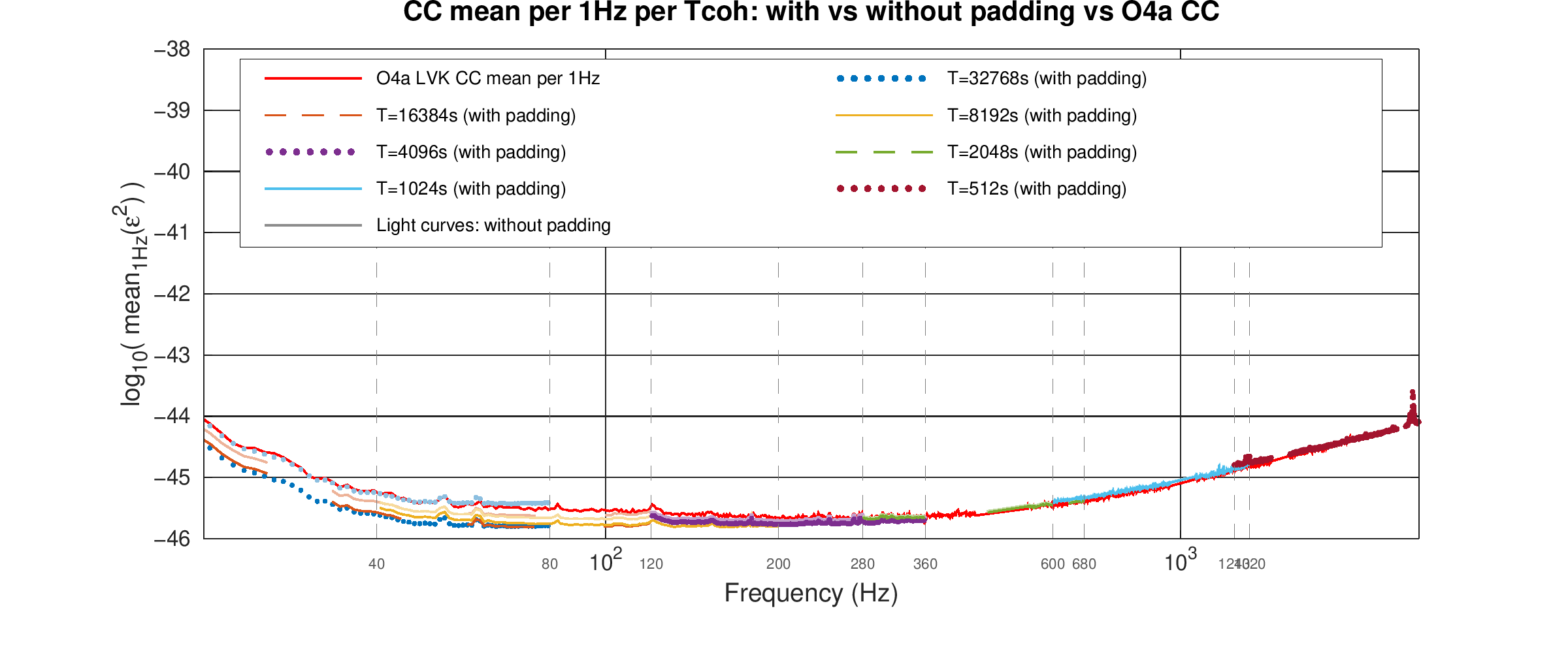}
\vspace{-0.1cm}

{\small (a) Cross-correlation\par}

\vspace{0.9cm}

\includegraphics[width=1.1\textwidth,height=1\textheight,keepaspectratio]{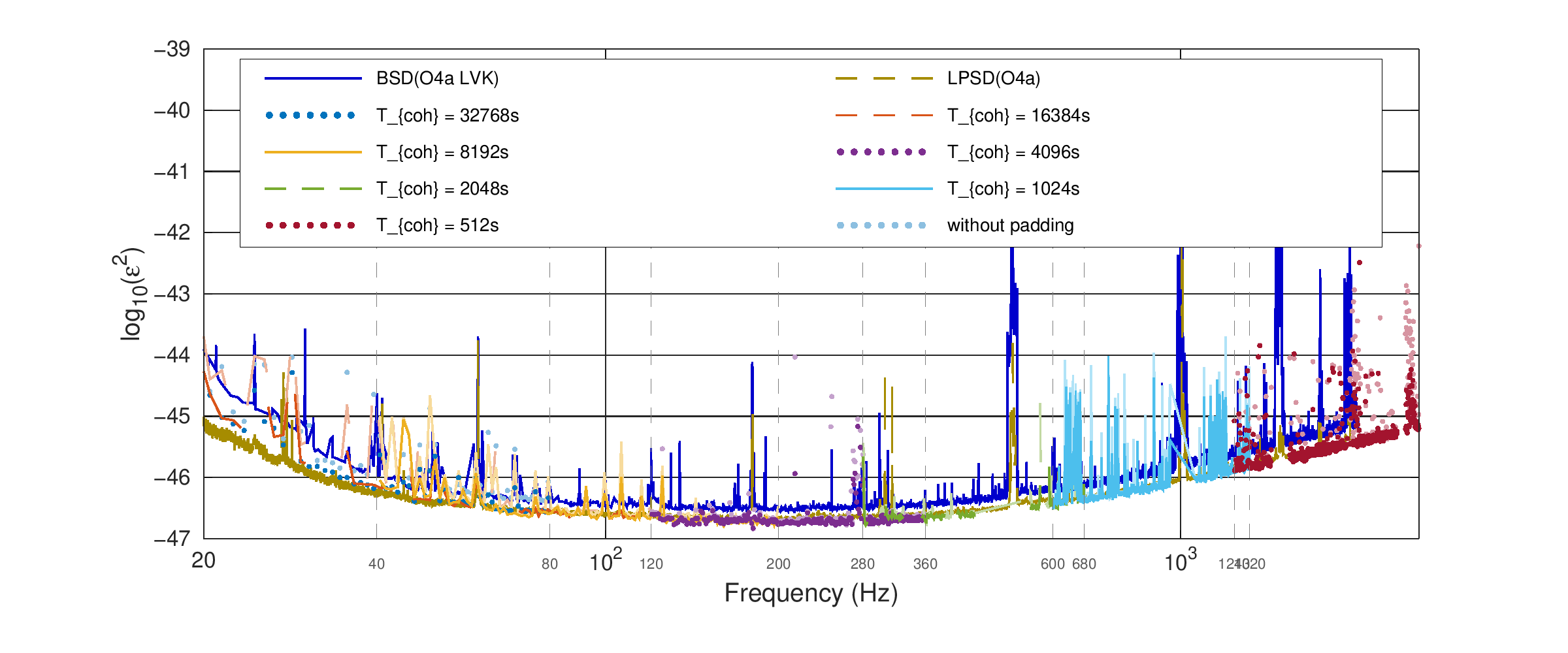}
\vspace{-0.1cm}

{\small (b) Combined H1--L1 semi-coherent\par}

\end{center}

\begingroup
\footnotesize
\refstepcounter{figure}
\label{fig:app_dpdm}
\noindent\textsc{Fig.~\thefigure.}\quad
Upper limits of the square of the dark-photon dark matter coupling strength,
$\varepsilon_{95\%}^{2}$, with different coherence times.
(a) The mean value of \textbf{Cross-correlation}.
(b) Max per Hz of \textbf{Combined H1--L1 semi-coherent}.
The curves in the overlap regions between adjacent coherence time bands are shown separately.
\par
\endgroup

\clearpage

\begin{center}

\includegraphics[width=1.1\textwidth,height=1\textheight,keepaspectratio]{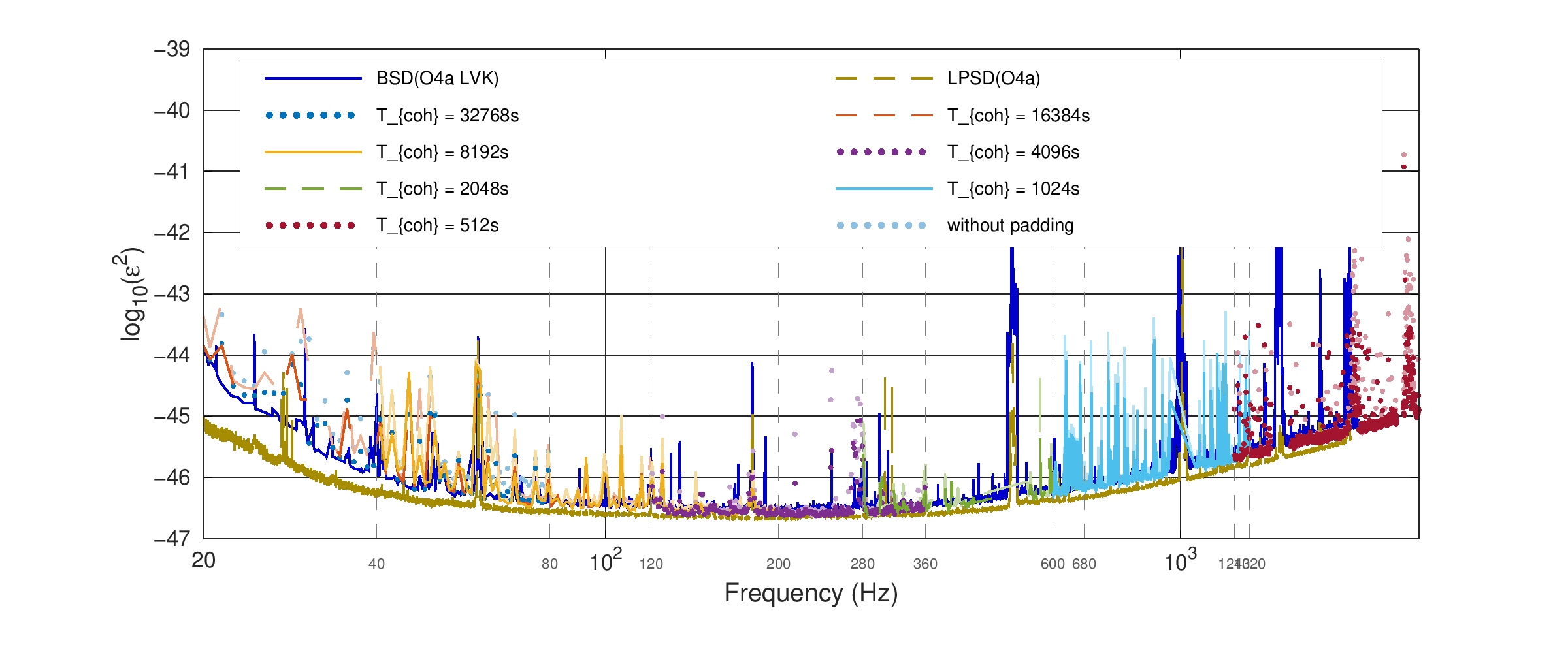}
\vspace{-0.1cm}

{\small (a) H1 semi-coherent\par}

\vspace{0.9cm}

\includegraphics[width=1.1\textwidth,height=1\textheight,keepaspectratio]{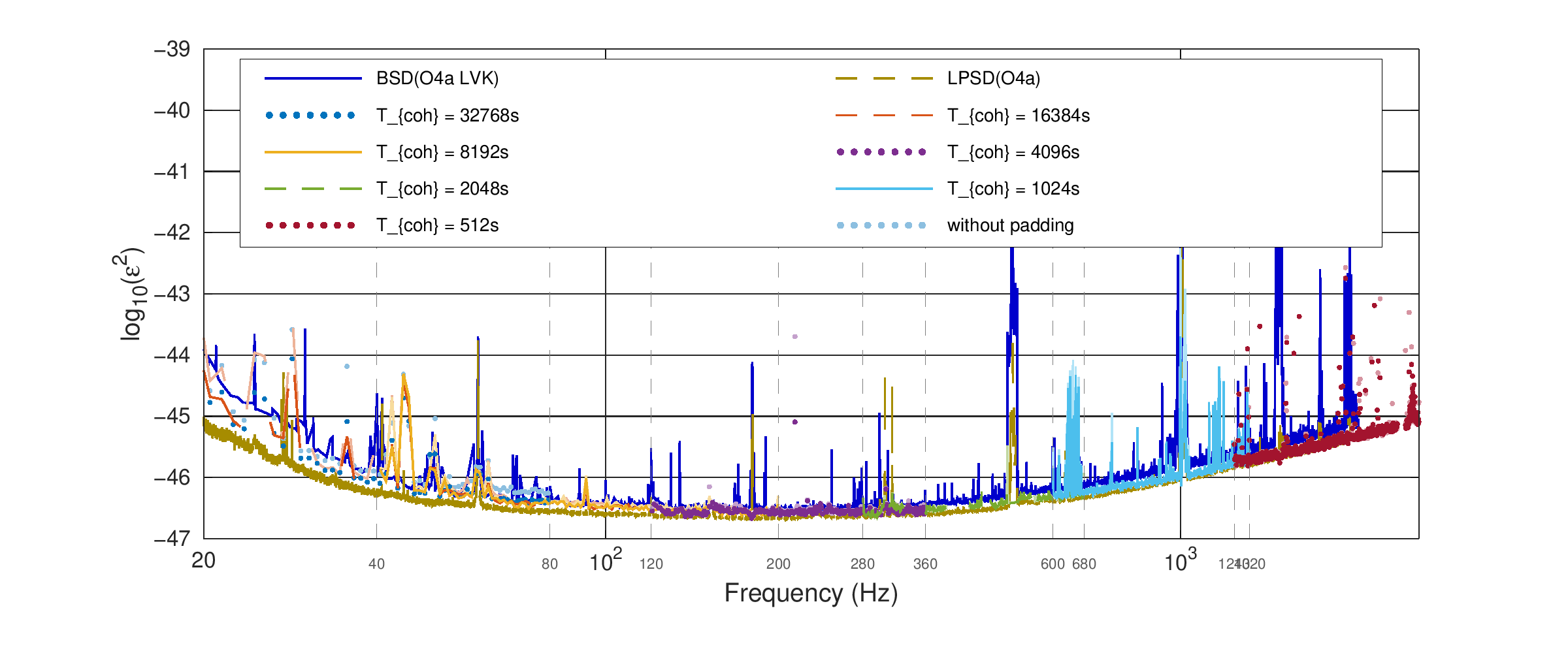}
\vspace{-0.1cm}

{\small (b) L1 semi-coherent\par}

\end{center}

\begingroup
\footnotesize
\refstepcounter{figure}
\label{fig:app_dpdm_h1_l1}
\noindent\textsc{Fig.~\thefigure.}\quad
Upper limits of the square of the dark-photon dark matter coupling strength,
$\varepsilon_{95\%}^{2}$, with different coherence times.
(a) \textbf{H1 semi-coherent}.
(b) \textbf{L1 semi-coherent}.
The curves in the overlap regions between adjacent coherence-time bands
are shown separately.
\par
\endgroup

\clearpage
\twocolumngrid
%%%%%%%%%%%%%%%%%%%%%%%%%%%%%%%%%%%%%%%%%%%%%%%%%
\section{Optimal weighting in the presence of zero padding}
\label{app:optimal_weighting}
In the following derivation, we suppress the detector index and denote the
livetime fraction of segment \(i\) by \(w_i\). Zero padding reduces the
coherent signal amplitude by \(w_i\) and the corresponding signal power by
\(w_i^2\). The measured zero-padded power can therefore be written as
\begin{equation}
d_i^M=w_i^2h_0^2+N_i^M,
\end{equation}
where \(h_0^2\) denotes the signal power for a fully live segment and \(N_i^M\) is the noise contribution with mean \(\overline N_i^M=\langle N_i^M\rangle\), while \(W_i\) denotes the statistical weight, which is to be optimized below. Correcting each measured power for the reduced signal response, the weighted statistic is
\begin{equation}
H=
\frac{
\displaystyle\sum_i W_i
\frac{d_i^M-\overline{N}_i^M}{w_i^2}
}{
\displaystyle\sum_i W_i
}.
\label{eq:app_general_stat}
\end{equation}

Using \(d_i^M=w_i^2h_0^2+N_i^M\) and
\(\langle N_i^M\rangle=\overline{N}_i^M\), the expectation value is
\begin{align}
\langle H\rangle
&=
\frac{
\displaystyle\sum_i W_i
\frac{
\left\langle d_i^M-\overline{N}_i^M\right\rangle
}{
w_i^2
}
}{
\displaystyle\sum_i W_i
}
\nonumber\\
&=
\frac{
\displaystyle\sum_i W_i
\frac{
w_i^2h_0^2+\langle N_i^M\rangle-\overline{N}_i^M
}{
w_i^2
}
}{
\displaystyle\sum_i W_i
}
=
h_0^2.
\label{eq:app_general_mean}
\end{align}
For statistically independent noise between different segments, the variance has the form
\begin{equation}
\operatorname{Var}(H)
=
\frac{
\displaystyle\sum_i W_i^2
\frac{\operatorname{Var}(N_i^M)}{w_i^4}
}{
\left(\displaystyle\sum_i W_i\right)^2
}.
\label{eq:app_general_var}
\end{equation}

Therefore, by minimizing Eq.~\eqref{eq:app_general_var} with respect to an individual weight \(W_k\),
\begin{equation}
\begin{aligned}
&\frac{\partial\,\operatorname{Var}(H)}{\partial W_k} \\
&=
\frac{2}{\left(\sum_iW_i\right)^3}
\left[
\frac{W_k\operatorname{Var}(N_k^M)}{w_k^4}\sum_iW_i
-
\sum_i\frac{W_i^2\operatorname{Var}(N_i^M)}{w_i^4}
\right]\\
&=0.
\end{aligned}
\label{eq:app_general_min}
\end{equation}
Since a common multiplicative factor in all \(W_i\) cancels in Eq.~\eqref{eq:app_general_stat}, we choose the optimal weight to be the form
\begin{equation}
W_i^{\mathrm{opt}}=\frac{w_i^4}{\operatorname{Var}(N_i^M)}.
\label{eq:app_general_weight}
\end{equation}

\subsection{Cross-correlation weighting}

For a simultaneous H1--L1 segment pair, the measured cross-power is \(\tilde{D}_{H,ij}\tilde{D}_{L,ij}^{*}\) and zero-padding reduces its signal response by a factor of \(w_{H,i}w_{L,i}\). For uncorrelated Gaussian detector noise,
\begin{equation}
\left\langle\tilde{D}_{H,ij}\tilde{D}_{L,ij}^{*}\right\rangle=0,
\end{equation}
and
\begin{equation}
\left\langle\left|\tilde{D}_{H,ij}\tilde{D}_{L,ij}^{*}\right|^2\right\rangle
=
\left\langle|\tilde{D}_{H,ij}|^2\right\rangle
\left\langle|\tilde{D}_{L,ij}|^2\right\rangle
=
P^M_{H,ij}P^M_{L,ij}.
\end{equation}
Since the real and imaginary parts have equal variance, we have
\begin{equation}
\operatorname{Var}\!\left[\operatorname{Re}\!\left(\tilde{D}_{H,ij}\tilde{D}_{L,ij}^{*}\right)\right]
=
\operatorname{Var}\!\left[\operatorname{Im}\!\left(\tilde{D}_{H,ij}\tilde{D}_{L,ij}^{*}\right)\right]
=
\frac{1}{2}P^M_{H,ij}P^M_{L,ij}.
\label{eq:app_cc_var}
\end{equation}
Applying Eq.~\eqref{eq:app_general_weight} to the cross-power and using
Eq.~\eqref{eq:app_cc_var} gives
\begin{equation}
W_{{\rm CC},i}^{\mathrm{opt}}
=
\frac{2(w_{H,i}w_{L,i})^2}{P^M_{H,ij}P^M_{L,ij}}.
\label{eq:app_cc_weight_pm}
\end{equation}

\subsection{Semi-coherent weighting}

For detector \(X\), the measured single-SFT power is \(|\tilde{D}_{X,ij}|^2\) and for circular complex Gaussian noise,
\begin{equation}
\left\langle|\tilde{D}_{X,ij}|^2\right\rangle=P^M_{X,ij},
\qquad
\left\langle|\tilde{D}_{X,ij}|^4\right\rangle=2\left(P^M_{X,ij}\right)^2.
\end{equation}
It follows that
\begin{equation}
\operatorname{Var}\!\left(|\tilde{D}_{X,ij}|^2\right)
=
\left\langle|\tilde{D}_{X,ij}|^4\right\rangle
-
\left\langle|\tilde{D}_{X,ij}|^2\right\rangle^2
=
\left(P^M_{X,ij}\right)^2.
\label{eq:app_pf_var}
\end{equation}

Equation~\eqref{eq:app_general_weight} therefore gives
\begin{equation}
W_{X,i}^{\mathrm{opt}}
=
\frac{w_{X,i}^4}{\left(P^M_{X,ij}\right)^2}.
\label{eq:app_pf_weight_pm}
\end{equation}
Notice that, although the general derivation above is written for the noise-subtracted
power, subtracting the mean noise contribution does not change its variance.
Therefore, the same optimal weight applies to the un-subtracted SFT powers
used in Eq.~\eqref{eq:pf_stat}, for which the local noise expectation is
removed subsequently in the SNR construction.
The weighting used here to account for zero-padding loss of signal
is analogous to that used in the PowerFlux continuous wave search algorithm~\cite{Abbott_2008}
to account for antenna pattern suppression of signal.
\clearpage
%%%%%%%%%%%%%%%%%%%%%%%%%%%%%%%%%%%%%%%%%%%%%%%%%
\bibliographystyle{apsrev4-2}
\bibliography{references}

\end{document}